\documentclass[11pt]{article}
\usepackage[T1]{fontenc}
\usepackage{amssymb}
\usepackage{amsfonts}
\usepackage{graphicx}
\usepackage{amsmath}

\usepackage{hyperref}

\makeatletter \@addtoreset{equation}{section} \makeatother

\newtheorem{theorem}{Theorem}
\newtheorem{lemma}{Lemma}

\newtheorem{remark}{Remark}

\newtheorem{proposition}{Proposition}

\begin{document}

\title{Central limit theorem for fluctuations of linear eigenvalue statistics of
large random graphs. Diluted regime.}
\author{ M. Shcherbina$^\dagger$
 \qquad B. Tirozzi$^*$\\
$^\dagger$Institute for Low Temperature Physics, Ukr. Ac. Sci \\
$^*$Department of Physics, Rome University "La Sapienza"
 }
\date{}

\maketitle

\begin{abstract}
We study the linear eigenvalue statistics of large random graphs in the regimes when the mean number of edges for each
vertex tends to infinity. We prove that for a rather wide class of test functions  the fluctuations of linear
eigenvalue statistics converges in distribution to a Gaussian random variable with zero mean and  variance which
coincides with "non gaussian" part of the Wigner ensemble variance.

\end{abstract}

\section{Introduction}\label{s:1}

In this paper we study the spectral properties of  ensembles of adjacency matrices of large random graphs.
 Following  Erd\H{o}s (see, e.g. \cite{JLR}), we introduce the probability measure  considering  the set of all graphs with
 $n$ vertices and set the weight of each graph $G$  as
\begin{equation}\label{1.1}
P(G) = (p_n/n)^{e(G)} (1-p_n/n)^{{n \choose 2} - e(G)},
\end{equation}
where $e(G)$ is the number of edges of $G$ and $0\le p_n\le n$. The set of $n$-vertices graphs with this measure
(usually denoted by ${\mathbf G}(n,p_n/n)$) is one of the classes of the prime reference in the theory of random
graphs. Most of the random graphs studies are devoted to the cases where $p_n/n\to 0$, as $n\to\infty$. There are two
major asymptotic regimes: $p_n \gg 1$ and $p_n = O(1)$ and corresponding models can be called {\it dilute random
graphs} and {\it sparse random graphs}, respectively.

It is well known that there is one-to-one correspondence between the graphs and their adjacency matrices. For ${\mathbf
G}(n,p_n/n)$ the ensemble corresponding   to (\ref{1.1}) consists of random symmetric $n\times n$ adjacency matrices $\widetilde
A$ is $\widetilde {\mathcal{A}}=\{\widetilde a_{ij}\}_{i,j=1}^n$ with $\widetilde a_{ii}=0$,
 and  i.i.d.
  \begin{equation}\label{ti-A}
\widetilde a_{ij}\!=\! \left\{ \begin{array}{ll} 1,&
\textrm{with} \ \textrm{probability } \ p_n/n ,
\\0,& \textrm{with} \ \textrm{probability} \ 1-p_n/n ,\\ \end{array}
\right.
\end{equation}
This is a  particular  case of the random matrix ensemble. Since the pioneering works by  Wigner \cite{W}  a big part
of the random matrix theory is devoted to the limiting transition $n\to\infty$.  The results obtained with this
limiting transition provide a rather good approximation of the spectral properties of random matrices (or random
graphs)  of a finite dimensionality.

An important advantage  of random matrices (\ref{ti-A}) is that their entries are independent up to the symmetry
 condition ($a_{ij}=a_{ji}$). This allows one to use the methods of random matrix theory which were developed to study
the classical Wigner matrices. Spectral properties of random adjacency matrix (\ref{A}) were examined in the limit
$n\to\infty$ both in numerical and theoretical physics studies.  The first results on the  spectral properties of
sparse and dilute random matrices in the physical literature are related with the works \cite{RB:88}, \cite{RD:90},
\cite{MF:91}, where equations for the limiting density of states of sparse random matrices were derived. In the papers
\cite{MF:91} and \cite{FM:96} a number of important  results on the universality of the correlation functions and the
Anderson localization transition were obtained.
  Unfortunately, these results were obtained with   non rigorous  replica and
super symmetry methods.

The first result on mathematical level of rigor for the matrices (\ref{ti-A}) was obtained in  \cite{BG2}, where the
eigenvalue distribution moments of the  matrix (\ref{ti-A})  with $p_n = p $ were studied in the limit $n\to\infty$. It
was shown that for any fixed natural $m$ there exists nonrandom limiting moment $\lim_{n\to\infty}n^{-1}\mathrm{Tr\,}
A^m$ and these moments can be found from the system of certain recurrent relations. The results of \cite{BG2} were
generalized to the case of weighted random graphs in \cite{KSV},  where the resolvent of the adjacency matrix was
studied and equations for the Stieltjes transform $g(z)$ of the limiting eigenvalue
 distribution were derived rigorously (note, that the  same equation for gaussian weights were obtained  in
  \cite{RB:88}, \cite{RD:90}, \cite{MF:91} by using the replica and
  the super symmetry approaches). But the limiting eigenvalue distribution, which is an analog of the low of large numbers
  of the probability theory, is only the first step in studies of  linear eigenvalue statistics, corresponding
  to the test function $\varphi$
 \begin{equation}\label{linst}
   \mathcal{N}_n[\varphi]=
\sum\varphi(\lambda_i)=\mathrm{Tr\,}\varphi(\mathcal{A}).
\end{equation}
Here and below $\{\lambda_{i}\}_{i=1}^n$ are eigenvalues of the matrix $\mathcal{A}$. The next step is to study the
behavior of fluctuations of linear eigenvalue statistics. For the case
  of sparse random matrices this step was done in \cite{ST:10} with some modification of  the method of
  \cite{KSV}.  It was shown in \cite{ST:10} that the random variable $n^{-1/2}(\mathcal{N}_n[\varphi]-
\mathbf{E}\{\mathcal{N}_n[\varphi]\})$  converges in  distribution  to the gaussian random variable, as $n\to\infty$
(here and below $\mathbf{E}\{...\}$ means the averaging with respect to all $\{a_{ij}\}_{1\le i<j\le n}$).

 The case of diluted  matrices ($p_n\to \infty$) is less complicated technically than that with $p_n=p$.
 It was shown in \cite{KKPS} that
 in this case to have  finite limits for $\mathbf{E}\{\mathcal{N}_n[\varphi]\}$
one should consider the matrix $\mathcal{A}'=\widetilde {\mathcal{A}}/\sqrt{p_n}$. Then it was proven in \cite{KKPS}
that for integrable test functions $\varphi$
  \[\lim_{n,p_n\to\infty,p_n/n\to 0}\mathbf{E}\{\mathcal{N}_n[\varphi]\}=\frac{1}{2\pi}\int_{-2}^{2}\varphi(\lambda)
  \sqrt{4-\lambda^2} d\lambda,\]
which coincides with the limits for the Wigner model \cite{W}. Let us note that the method, used in \cite{KKPS}, is
rather similar to that for the Wigner model. But the problem  to study the fluctuations of linear eigenvalue statistics
usually is much more complicated than the problem to find the limiting eigenvalue distribution of random matrix
ensemble. Even for the classical Wigner case the central limit theorem (CLT) for  fluctuations of linear eigenvalue
statistics was proven only recently in the series  of  papers with  improving results
\cite{Si-So:98a,B:04,LP:09,S:10}.

In the present paper we prove CLT for  fluctuations of linear  eigenvalue statistics of diluted matrices, more
precisely, we prove that the random variable $(p_n/n)^{1/2}(\mathcal{N}_n[\varphi]-
\mathbf{E}\{\mathcal{N}_n[\varphi]\})$ in the limit $n,p_n\to\infty,p_n/n\to 0$ converges in  distribution to the
normal random variable. The method of the paper is a generalization of that of \cite{S:10}. It allows us to prove CLT
under rather weak assumptions on the test function $\varphi$ (see Theorems \ref{t:CLT} and \ref{t:CLT2} below).

  It will be  more convenient for us to study the
matrix $\mathcal{A}=\widetilde{ \mathcal{A}}/\sqrt{p_n}-\mathbf{E}\{\widetilde{ \mathcal{A}}/\sqrt{p_n}\}$, where
$\mathbf{E}\{...\}$ means averaging with respect to all entries of $\widetilde{\mathcal{A}}$. It is easy to see that
$\mathcal{A}$ differs from $\mathcal{A}'$ by the rank one matrix $\mathbf{E}\{\widetilde{ \mathcal{A}}/\sqrt{p_n}\}$.
So, everywhere below we will assume that the entries $a_{ij}$ of $\mathcal{A}$ are distributed as
\begin{equation}\label{A}
 a_{ij}\!=\! \left\{ \begin{array}{ll} \dfrac{1}{\sqrt{p_n}}-\dfrac{\sqrt{p_n}}{n},&
\textrm{with} \ \textrm{probability } \ p_n/n ,
\\-\dfrac{\sqrt{p_n}}{n},& \textrm{with} \ \textrm{probability} \ 1-p_n/n ,\\ \end{array}
\right.
\end{equation}
Let us note that the case $p_n\sim\alpha n$ here corresponds to the Wigner ensemble, hence the model (\ref{A}) allows
us to make "smooth transition" from the matrix studied in \cite{KSV} to the Wigner matrix.

Let us set our main notations.
 For any measurable function $f$ we  denote by $\mathbf{E}\{f(\mathcal{A})\}$ the averaging
with respect to all random variables $\{a_{ij}\}_{1\le i<j\le n}$ and
\begin{equation}\label{dVar}
\mathbf{Var}\{f(\mathcal{A})\}:=\mathbf{E}\{|f(\mathcal{A})-\mathbf{E}\{f(\mathcal{A})\}|^2\}.
\end{equation}
We denote also for any random variable $\xi$
\[ \overset\circ{\xi}=\xi^\circ=\xi-\mathbf{E}\{\xi\}.\]
Introduce the resolvent of  $\mathcal{A}$
\begin{equation}\label{G}
G_{jk}(z)=(\mathcal{A}-z)^{-1}_{jk},\quad\Im z\not=0, \quad\gamma_n(z)=\hbox{Tr\,}G(z).
\end{equation}
In what follows it will be important for us that
\begin{align}\label{f<1}
&||G||\le |\Im z|^{-1},\quad\sum_{j=1}^n|G_{ij}|^2=(GG^*)_{ii}\le||G||^2\le |\Im z|^{-2},\\
&\Im (Ge,e)\Im z\ge 0,\quad \;\forall e\in\mathbb{R}^n.
\end{align}
Here and everywhere below $||\mathcal{A}||$ means the operator norm of the matrix $\mathcal{A}$.

The main result of the paper is  the central limit theorem for
the linear eigenvalue statistics of any sufficiently smooth function $\varphi$ which  grows not
faster than exponential at infinity. But we prove  CLT first for the functions,
which are smooth enough and decaying. Set
\begin{equation}\label{norm}
    ||\varphi||_s^2=\int(1+2|k|)^{2s}|\widehat\varphi(k)|^2dk,\quad
    \widehat\varphi(k)=\frac{1}{2\pi}\int e^{ikx}\varphi(x)dx
\end{equation}
and let $\mathcal{H}_s$  be the space of all function possessing the norm $||.||_s$.
\begin{theorem}\label{t:CLT}
Consider the adjacency matrix (\ref{A}) with $p_n\to\infty, p_n/n\to 0$.
 Assume that the real valued function $\varphi\in\mathcal{H}_s$ with $s>3/2$ and  that
\begin{equation}\label{cond}
\int_{-2}^{2}\varphi (\mu )\frac{2-\mu ^{2}}{\sqrt{4-\mu ^{2}}}d\mu \not=0.
\end{equation}
  Then
the random variable $(p_n/n)^{1/2}\overset\circ{\mathcal{N}_n}[\varphi]$ converges in distribution to  a Gaussian
random variable with zero mean and  variance
\begin{equation}
V[\varphi ]=\frac{1}{2\pi ^{2}}\left(
\int_{-2}^{2}\varphi (\mu )\frac{2-\mu ^{2}}{\sqrt{4-\mu ^{2}}}%
d\mu \right) ^{2}.  \label{limvar}
\end{equation}%
\end{theorem}

It is interesting to compare (\ref{limvar}) with that for the Wigner model
\[M=n^{-1/2}\{w_{ij}\}_{i,j=1}^n,\quad E\{w_{ij}\}=0,\quad E\{|w_{ij}|^2\}=1,\quad (i\not=j),\]
we have (see \cite{S:10})
\begin{eqnarray*}
V_W[\varphi]=\lim_{n\to\infty}\mathbf{Var}\{\overset\circ{\mathcal{N}}_n[\varphi ]\}=\frac{1}{2 \pi
^{2}}\int_{-2}^{2}\int_{-2}^{2}\left( \frac{ \varphi(\lambda_1)-\varphi(\lambda_2)}{\lambda_1-\lambda_2 }\right)
^{2}\frac{(4-\lambda _{1} \lambda _{2})d\lambda _{1}d\lambda _{2}}{\sqrt{4-\lambda
_{1}^{2}}\sqrt{4-\lambda _{2}^{2}}}\\
+\frac{\kappa _{4}}{2\pi ^{2}}\left(
\int_{-2}^{2}\varphi (\mu )\frac{2-\mu ^{2}}{\sqrt{4-\mu ^{2}}}%
d\mu \right) ^{2}+\frac{w_2-2}{4\pi ^{2}}\left(
\int_{-2}^{2}\frac{\varphi (\mu )\mu d\mu }{\sqrt{4-\mu ^{2}}}%
\right) ^{2},
\end{eqnarray*}
Here $\kappa _{4}=n^2(\mathbf{E}\{M_{ij}^4\}-3\mathbf{E}^2\{M_{ij}^2\})=\mathbf{E}\{w_{ij}^4\}-3$,
$w_2=nE\{|M_{ii}|^2\}$. One can see that (\ref{limvar}) coincides with the term multiplying $\kappa _{4}$. This can be
understood if we recall that in our case $\kappa_4=n^2(\mathbf{E}\{a_{ij}^4\}-3\mathbf{E}^2\{a_{ij}^2\})\sim n/p_n$ and
we consider the random variable $(p_n/n)^{1/2}\overset\circ{\mathcal{N}_n}[\varphi]$, while in the Wigner case one
should consider $\overset\circ{\mathcal{N}_n}[\varphi]$.

One more interesting question is what  is happening if the l.h.s. of (\ref{cond}) is zero. It is easy to guess that in
this case one have to change the normalization factor in front of $\overset\circ{\mathcal{N}_n}[\varphi]$. But it could
happen that the new expression for the limiting variance in this case will depend on the rate of convergence of
$p_n/n\to 0$. We are going to study this situation in  the future works.

 Consider the set $\mathcal{H}_s^{(c)}$ of the functions, represented in the form
\begin{equation}\label{H_s^c}
    \varphi(\lambda)=\cosh
    (c\lambda)\,\widetilde\varphi(\lambda),\quad\widetilde\varphi\in\mathcal{H}_s.
\end{equation}
\begin{theorem}\label{t:CLT2} Consider the adjacency matrix (\ref{A}) with $p_n\to\infty, p_n/n\to 0$.
 Assume that the real valued function $\varphi\in\mathcal{H}_s^{(c)}$    with some $c>0$, $s>3/2$
 and (\ref{cond}) is satisfied. Then
the random variable
$(p_n/n)^{1/2}\overset\circ{\mathcal{N}_n}[\varphi]$ converges in distribution to  a Gaussian
random variable with zero mean and  variance (\ref{limvar}).
\end{theorem}

\section{Proofs}

The proof follows the strategy developed in cite{S:10} for the Wigner model. We start from the  lemma

\begin{lemma}\label{l:1} Let $\gamma_n(z)$ be defined by (\ref{G}). Then for
 any  $z:\Im z>0$   there exists a constant $C$ such that
  \begin{align}\label{b_var}
\frac{p_n}{n}\mathbf{Var}\{\gamma_n(z)\}\le  C/|\Im z|^4,\quad
\left(\frac{p_n}{n}\right)^2\mathbf{E}\{|\gamma_n^{\circ}(z)|^4\}\le C/|\Im z|^{12}.
\end{align}
Moreover, for any $\varepsilon>0$ we have
\begin{align}\label{b_varm}
\frac{p_n}{n}\mathbf{Var}\{\gamma_n(z)\}\le C\mathbf{E}\{|G_{11}|^{1+\varepsilon}\}/|\Im z|^{3+\varepsilon},
\end{align}
and for any smooth function  $F$ and any $z:\Im z>a$
\begin{equation}\label{l1.1}
\mathbf{Var}\Big\{n^{-1}\sum_{j=1}^nF(G_{jj}(z))\Big\}\le n^{-1}\sup_{\zeta:0<\Im \zeta,
|\zeta|<a^{-1}}|F'(\zeta)|^2.
\end{equation}
\end{lemma}

\textit{Proof of Lemma \ref{l:1}}
To prove (\ref{b_var}) we use the following proposition proven in \cite{Dh-Co:68}
 \begin{proposition}
\label{p:mart} Let $\xi _{\alpha },\;\alpha =1,...,\nu $ be independent
random variables, assuming values in $\mathbb{R}^{m_{\alpha }}$ and having
probability laws $P_{\alpha }$, $\alpha =1,\dots ,\nu $ and let $\Phi :%
\mathbb{R}^{m_{1}}\times \dots \times \mathbb{R}^{m_{\nu }}\rightarrow
\mathbb{C}$ be a Borelian function. Set
\begin{equation}\label{phi_j}
\Phi _{\alpha }(\xi _{1},\dots ,\xi _{\alpha })=\int \Phi (\xi _{1},\dots
,\xi _{\alpha },\xi _{\alpha +1},\dots ,\xi _{\nu })P_{\alpha +1}(d\xi
_{\alpha +1})\dots P_{\nu }(d\xi _{\nu })
\end{equation}%
so that $\Phi _{\nu }=\Phi ,\quad \Phi _{0}=\mathbf{E}\{\Phi \}$,
where $\mathbf{E}\{\dots \}$ denotes the expectation with respect to the
product measure $P_{1}\dots P_{\nu }$.

Then for any positive $p\geq 1$ there exists $C_{p}^{\prime }$, independent
of $\nu $ and such that
\begin{equation}
\mathbf{E}\{|\Phi -\mathbf{E}\{\Phi \}|^{2p}\}\leq C_{p}^{\prime }\nu
^{p-1}\sum_{\alpha =1}^{\nu }\mathbf{E}\{|\Phi _{\alpha }-\Phi _{\alpha
-1}|^{2p}\}.\label{martk}
\end{equation}
\end{proposition}
 Let $\Phi=\gamma_n(z)$, $\xi_{\alpha}=\{a_{\alpha j}\}_{j\le\alpha}$. Denote also $E_\alpha\{.\}$
 the averaging with respect to the random variables $\{a_{\alpha j}\}_{j=1}^{n}$. Then
it is easy to see that
\[\Phi _{\alpha }=E_{\alpha+1}\dots E_{n}\Phi\]
and by the H\"{o}lder inequality
\begin{equation}\mathbf{E}\{|\Phi _{\alpha }-\Phi _{\alpha
-1}|^{2p}\}=\mathbf{E}\{|E_{\alpha+1}\dots E_{n}(\Phi -E_\alpha\{\Phi)\}|^{2p}\}
 \le \mathbf{E}\{|\Phi-\mathbf{E}_\alpha\{\Phi\}|^{2p}\}=\mathbf{E}\{|\Phi-\mathbf{E}_1\{\Phi\}|^{2p}\}\label{martk1}
\end{equation}

Define $\mathcal{A}^{(1)}$ as a $(n-1)\times (n-1)$ matrix which can be obtained from $\mathcal{A}$ if we remove from
$\mathcal{A}$ the first line and the first column. Set also
\begin{align}\label{A^1}
 G^{(1)}(z)&=(\mathcal{A}^{(1)}-z)^{-1},
\quad\gamma_n^{(1)}(z)=\sum_{i=2}^nG^{(1)}_{ii}(z),\quad a^{(1)}=(a_{12},\dots,a_{1n}).
\end{align}
We use the representations:
\begin{align}\label{rep_1}
G_{11}(z)&=-(z+
(G^{(1)}a^{(1)},a^{(1)}))^{-1},\\
G_{ii}(z)&=G_{ii}^{(1)}(z)-\frac{(G^{(1)}a^{(1)})_i(G^{(1)}a^{(1)})_i}{z+ (G^{(1)}a^{(1)},a^{(1)})},\quad i\not=1.
\notag\end{align} Since $G^{(1)}$ does not depend on $a^{(1)}$, we have
\begin{align}\notag
\gamma_n-\mathbf{E}_1\{\gamma_n\}&=-\frac{1+(G^{(1)} G^{(1)}a^{(1)},a^{(1)})}{z+
(G^{(1)}a^{(1)},a^{(1)})}+\mathbf{E}_1\bigg\{\frac{1+(G^{(1)} G^{(1)}a^{(1)},a^{(1)})}{z+
(G^{(1)}a^{(1)},a^{(1)})}\bigg\}\\
&:=-\frac{1+B(z)}{A(z)}+\mathbf{E}_1\Big\{\frac{1+B(z)}{A(z)}\Big\}. \label{repr}\end{align} Hence,  it suffices to
estimate $\mathbf{E}\{|B/A-\mathbf{E}_{1}\{B/A\}|^2\}$ and $\mathbf{E}\{|A^{-1}-\mathbf{E}_{1}\{A^{-1}\}|^2\}$. We show
how to estimate the first expression. The second one can be estimated similarly. Denote by
$\xi^\circ_1=\xi-\mathbf{E}_{1}\{\xi\}$ for any random variable $\xi$. Note that since for any $a$
\begin{equation}\label{in_a}
\mathbf{E}_{1}\{|\xi-a|^2\}=\mathbf{E}_{1}\{|\xi^\circ_1|^2\}+|a-\mathbf{E}_{1}\{\xi\}|^2\Rightarrow
\mathbf{E}_{1}\{|\xi^\circ_1|^2\}\le\mathbf{E}_{1}\{|\xi-a|^2\}\end{equation}
 it suffices to estimate
$\mathbf{E}\{|B/A-\mathbf{E}_{1}\{B\}/\mathbf{E}_{1}\{A\}|^2\}$ instead $\mathbf{E}\{|B/A-\mathbf{E}_{1}\{B/A\}|^2\}$.
Then it is easy to see that
\begin{equation}\label{in_a1}\bigg|\frac{B}{A}-\frac{\mathbf{E}_{1}\{B\}}{\mathbf{E}_{1}\{A\}}\bigg|=
\bigg|\frac{B^\circ_1}{\mathbf{E}_{1}\{A\}}- \frac{A^\circ_1}{\mathbf{E}_{1}\{A\}}\,\frac{B}{A}\bigg|\le
\bigg|\frac{B^\circ_1}{\mathbf{E}_{1}\{A\}}\bigg|+ \bigg|\frac{A^\circ_1}{\Im z\mathbf{E}_{1}\{A\}}\bigg|.\end{equation}

Here we  used  the relations that follow from the spectral theorem
\begin{align}\notag
\Im(G^{(1)}a^{(1)},a^{(1)})=\Im z(G^{(1)}a^{(1)},G^{(1)}a^{(1)}),\quad
\Im\mathrm{Tr\,} G^{(1)}=\Im z\mathrm{Tr\,} (G^{(1)}G^{(1)*})\\
\Rightarrow\frac{(G^{(1)}a^{(1)},G^{(1)}a^{(1)})}{|z+(G^{(1)}a^{(1)},a^{(1)})|}\le|\Im z|^{-1},\quad
\frac{n^{-1}\mathrm{Tr\,} (G^{(1)}G^{(1)*})}{|z+n^{-1}\mathrm{Tr\,} G^{(1)}|}\le|\Im z|^{-1}. \label{sp_rel}\end{align}
The first relation yields, in particular, that $|{B}/{A}|\le|\Im z|^{-1}$. It is evident that
\begin{align}\label{A^0}
&&A^\circ_1=\sum_{i\not=j}G^{(1)}_{ij}a_{1i}a_{1j}+\sum_{i}G^{(1)}_{ii}(a_{1i}^2)^\circ,\\
&&\mathbf{E}_1\{|A^\circ_1|^2\}\le C(p_nn)^{-1}\mathrm{Tr\,} (G^{(1)}G^{(1)*}). \notag\end{align} In view of
(\ref{sp_rel}) and (\ref{f<1}) we have
\begin{equation}\label{*}\frac{n^{-1}\mathrm{Tr\,} (G^{(1)}G^{(1)*})}{|z+n^{-1}\mathrm{Tr\,} G^{(1)}|}\le
\frac{|\Im z|^{-1+\varepsilon}|n^{-1}\mathrm{Tr\,} (G^{(1)}G^{(1)*})|^\varepsilon} {|z+n^{-1}\mathrm{Tr\,}
G^{(1)}|^\varepsilon}\le C\frac{|\Im z|^{-1-\varepsilon}}{|\mathbf{E}_{1}\{A\}|^{\varepsilon}}.\end{equation}
 Here in the first
inequality  the numerator $N$ and the denominator $D$ are just written as $N=N^\varepsilon N^{1-\varepsilon},
D=D^\varepsilon D^{1-\varepsilon}$, then for $(N/D)^{1-\varepsilon}$ the second inequality of (\ref{sp_rel}) is used,
and then for $N^\varepsilon$ the inequality (\ref{f<1}) is used. Thus, in view of the second line of (\ref{A^0})
\[
\mathbf{E}_1\bigg\{\bigg|\frac{A^\circ_1}{\mathbf{E}_{1}\{A\}}\bigg|^2\bigg\}\le C\frac{(p_nn)^{-1}\mathrm{Tr\,}
(G^{(1)}G^{(1)*})}{|z+n^{-1}\mathrm{Tr\,} G^{(1)}|^2}\le C\frac{|\Im
z|^{-1-\varepsilon}}{p_n|\mathbf{E}_{1}\{A\}|^{1+\varepsilon}}.
\]

Similarly
\[\mathbf{E}_1\bigg\{\bigg|\frac{B^\circ_1}{\mathbf{E}_{1}\{B\}}\bigg|^2\bigg\}
\le C\frac{\mathrm{Tr\,} (G^{(1)}G^{(1)}G^{(1)*}G^{(1)*})}{p_nn|z+n^{-1}\mathrm{Tr\,} G^{(1)}|^2}\le
C\frac{p_n^{-1}n^{-1}\mathrm{Tr\,} (G^{(1)}G^{(1)*})}{|\Im z|^{2}|z+n^{-1}\mathrm{Tr\,} G^{(1)}|^2}\le
C\frac{p_n^{-1}|\Im z|^{-3-\varepsilon}}{|\mathbf{E}_{1}\{A\}|^{1+\varepsilon}},\] because, using the averaging with
respect to $\{a_{1j}\}$, we obtain for $E_1\{|B^\circ|^2\}$ the same bound as in the second line of (\ref{A^0}), but
with $G^{(1)}$ replaced by $(G^{(1)})^2$. This gives the first inequality above. Then we use that $\mathrm{Tr\,}
(G^{(1)}G^{(1)}G^{(1)*}G^{(1)*})\le|\Im z|^{-2}\mathrm{Tr\,} (G^{(1)}G^{(1)*})$ (since $||G^{(1)}||^2\le|\Im z|^{-2}$
and finally use (\ref{*}).

Then,   the Jensen inequality  $|\mathbf{E}_1\{A\}|^{-1}\le\mathbf{E}_1\{|A|^{-1}\}$,
and the relation $A^{-1}=-G_{11}(z)$
yield
\[\mathbf{E}\{|(\gamma_n(z))^\circ_1|^2\}\le \frac{C\mathbf{E}\{|G_{11}(z)|^{1+\varepsilon}\}}
{p_n|\Im z|^{3+\varepsilon}}.\]
Then (\ref{martk}) for $p=1$ implies (\ref{b_varm}). Putting here $\varepsilon=0$ we get
(\ref{b_var}).

To prove the second inequality of (\ref{b_var}), we use  (\ref{martk}) for $p=2$.
In view of  (\ref{repr}) it is enough to check that
\begin{equation}\label{b_v.4}
    \mathbf{E}_1\{|A^\circ_1|^4\}\le Cp_n^{-2}|\Im z|^{-4},\quad
\mathbf{E}_1\{|B^\circ_1|^4\}\le Cp_n^{-2}|\Im z|^{-8}.
\end{equation}
The first relation here evidently follow from (\ref{A^0}), if we take the fourth power and average
with respect to $\{a_{1i}\}$.  The second one can be obtained similarly.

To prove (\ref{l1.1}) we note first that (\ref{martk}) and (\ref{martk1}) for $\Phi=n^{-1}\sum
F(G_{jj})$ yield
\begin{align*}
\mathbf{Var}\{\Phi\}\le n\mathbf{E}\Big\{\Big|\Phi-\mathbf{E}_1\{\Phi\}\Big|^2\Big\}\le
n^{-1}\mathbf{E}\Big\{\Big|\sum_j \big(F(G_{jj})-F(G_{jj}^{(1)})\big)\Big|^2\Big\}\\
\le n^{-1}\sup_{\zeta:0<\Im \zeta, |\zeta|<a^{-1}}|F'(\zeta)|^2\mathbf{E}
\Big\{\Big(\sum_j \big|G_{jj}-G_{jj}^{(1)}\big|\Big)^2\Big\},
\end{align*}
if we take into account that $n^{-1}\sum F(G_{jj}^{(1)})$ does not depend on $a_{1j}$ and hence may
play the role of  $a$ in the inequality (\ref{in_a}). Moreover,
 using (\ref{rep_1}) and (\ref{sp_rel}), we get
\[\sum_j \big|G_{jj}-G_{jj}^{(1)}\big|\le \frac{1+(G^{(1)}a^{(1)},G^{(1)}a^{(1)})}{|z+G^{(1)}a^{(1)},a^{(1)})|}
\le|\Im z|^{-1}.\]
The above two bounds prove (\ref{l1.1}).
$\square$

Lemma \ref{l:1} gives the bound for the variance of the linear eigenvalue statistics
for the functions $\varphi(\lambda)=(\lambda-z)^{-1}$. Now we are going to extend the bound
to a  wider class of test functions. For this aim we use Proposition \label{p:joh} below.
We formulate it for the variance of linear eigenvalue statistics, but one can see easily that
it can be applied also to a more general case even without reference to random matrix, see
e.g. \cite{So-Co:11}.
Proposition \label{p:joh} was proven in \cite{S:10}, but for the completeness we give its proof here.
We also would like to thank
Prof. A.Soshnikov for the fruitful discussion on the proposition, which allows us to make proof
the proof more simple.

\begin{proposition}\label{p:joh} Let $\mathcal{A}$ be any random $n\times n$ matrix,  $\mathcal{N}_n[\varphi]$
be its linear eigenvalue statistic (\ref{linst}), and $\gamma_n(z)$ be defined by (\ref{G}). Then
\begin{equation}\label{pj.1}
\mathbf{Var}\{\mathcal{N}_n[\varphi]\}\le C_s||\varphi||_s^2\int_0^\infty dy
e^{-y}y^{2s-1}\int_{-\infty}^\infty\mathbf{Var}\{\gamma_n(x+iy)\}dx
\end{equation}
where $||\varphi||_s$ is defined in (\ref{norm}).
\end{proposition}
\begin{remark} If the integral in the r.h.s. is equal to infinity, then the inequality is not interesting,
hence we will assume that this integral is finite.
\end{remark}
\textit{Proof.}  Consider the operators $\mathcal{D}_s$, $\mathcal{V}$ defined in the space of the Fourier
transforms of the functions of the standard $L_2(\mathbb{R})$:
\begin{align}\notag
   &\widehat{ \mathcal{D}_sf}(k)=(1+2|k|)^s\widehat f(k),\\
   &\widehat{\mathcal{V}f}(k)=\int dk'
   \widehat{\mathcal{V}}(k,k')\widehat{f}(k'),\quad
   \widehat{\mathcal{V}}(k_1,k_2)=\mathbf{Cov}\{\hbox{Tr}e^{ik_1\mathcal{A}},\hbox{Tr}e^{ik_2\mathcal{A}}\}.
\label{D_s}\end{align}
It is easy to see that if we introduce the operator $K:=\mathcal{D}_s^{-1}
\mathcal{V}\mathcal{D}_s^{-1}$ then
\begin{align}\notag
\mathbf{Var}\{\mathcal{N}_n[\varphi]\}&=(2\pi)^{-2}(\mathcal{V}\varphi,\varphi)=(2\pi)^{-2}(K\mathcal{D}_s\varphi,\mathcal{D}_s\varphi)\\
&\le(2\pi)^{-2}||K||\cdot||\mathcal{D}_s\varphi||^2\le(2\pi)^{-2}||\varphi||_s^2\hbox{Tr} K
\label{calV}\end{align}

Let us check that the operator $K$  is indeed of the trace class in $L_2(\mathbb{R})$.
Note first that in the Fourier space his kernel has the form
\[\widehat K(k_1,k_2)=(1+2|k_1|)^{-s}\widehat{\mathcal{V}}(k_1,k_2)(1+2|k_1|)^{-s},\]
with $\widehat{\mathcal{V}}(k_1,k_2)$ of (\ref{D_s}).
It is evident that  $K\ge 0$ in the operator sense, and  $K(k_1,k_2)$ is a continuous function of $k_1,k_2$, since
 $\widehat{\mathcal{V}}(k_1,k_2)$ can be written as a finite sum of the products of  Fourier transforms
 of the positive unit measures, which are the  distributions of  eigenvalues of $\mathcal{A}$.  Moreover, we will prove below that
\begin{equation}\label{pj.2a} \int\widehat K(k,k)dk<\infty.
\end{equation}
Then it follows from the inequality $|\widehat K(k_1,k_2)|^2\le \widehat K(k_1,k_1)\widehat K(k_2,k_2)$ (which is valid for
any continuous positive definite kernels) that $\widehat K$ belongs to the Hilbert-Schmidt class, and therefore $\widehat K$ has
a basis $\{\phi_j(k)\}_{j=1}^\infty$, which is made from the continuous eigenfunctions
  with corresponding eigenvalues $\lambda_j>0$.  Then if we consider a finite rank operator with the kernel
  $\widehat K_N(k_1,k_2)=\sum_{j=1}^N\lambda_j\phi_j(k_1)\bar\phi_j(k_2)$, we have $ K- K_N\ge 0$
   in the operator sense, and
   $\widehat K(k_1,k_2)-\widehat K_N(k_1,k_2)$ is continuous, thus $\widehat K(k,k)\ge
   \widehat K_N(k,k)$ and
   \[\sum_{j=1}^N\lambda_j= \int \widehat K_N(k,k)dk\le\int \widehat K(k,k)dk.\]   Since  $N$ is arbitrary and $\lambda_j>0$
   we obtain that $K$ is a trace class operator.

We are left to prove the inequality of (\ref{pj.2a}).
We have
\begin{align*}
\int \widehat K(k,k)dk&=\int(1+2|k|)^{-2s}\widehat{\mathcal{V}}(k,k)dk\\
&=\frac{1}{\Gamma(2s)}\int_0^\infty dy
e^{-y}y^{2s-1}\int e^{-2|k|y}\widehat{\mathcal{V}}(k,k)dk\\
&=\frac{1}{\Gamma(2s)}\int_0^\infty dy
e^{-y}y^{2s-1}\int dx \int\int dk_1 dk_2 e^{i(k_1-k_2)x}\widehat{\mathcal{V}}(k_1,k_2)e^{-|k_1|y-|k_2|y}\\
&=\frac{1}{\Gamma(2s)}\int_0^\infty dy
e^{-y}y^{2s-1}\int dx\mathbf{Var}\{\mathcal{N}_n[P_y(x-.)]\}\\
&=\frac{1}{\Gamma(2s)}\int_0^\infty dy e^{-y}y^{2s-1}\int dx\mathbf{Var}\{\Im\gamma_n(x+iy)\},
\end{align*}
where  $P_y$ is the Poisson kernel
\begin{equation}\label{P_y}
    P_y(x)=\frac{y}{\pi(x^2+y^2)}.
\end{equation}
and we used that
\[\int P_y(x-\lambda)e^{ik\lambda}d\lambda=e^{ikx-|k|y}.\]
This relation combined with (\ref{calV}) proves (\ref{pj.1}).$\square$

\medskip

Now we are ready to prove the bound for the variance of linear eigenvalue statistics
for a rather wide class of the test functions
\begin{lemma}\label{l:2} If $||\varphi||_{3/2+\alpha}\le\infty$, with any $\alpha>0$, then
\begin{equation}\label{l2.1}
 \frac{p_n}{n}  \mathbf{ Var}\{\mathcal{N}_n[\varphi]\}\le C_\alpha||\varphi||_{3/2+\alpha}^2
\end{equation}
\end{lemma}
\textit{Proof.}
In view of Proposition
\ref{p:joh} we need to estimate
\[I(y)=\int_{-\infty}^{\infty}\mathbf{Var}\{\gamma_n(x+iy)\}dx
\]
Take in (\ref{b_varm})  $\varepsilon=\alpha/2$. Then  we need to estimate
\[\int_{-\infty}^{\infty}\mathbf{E}\{|G_{11}(x+iy)|^{1+\alpha/2}\}dx.
\]
Use the spectral representation
\[{G}_{11}=\int\frac{N_{11}(d\lambda)}{\lambda-x-iy},\quad\mathrm{were}\quad
N_{11}(\Delta)=\sum_{k=1}^n|\psi^{(k)}_1|^2\mathbf{1}_\Delta(\lambda_k)\]
with $\psi^{(k)}=(\psi^{(k)}_1,\dots,\psi^{(k)}_n)$ being an eigenvector of $\mathcal{A}$, corresponding the
eigenvalue $\lambda_k$, i.e.
$\mathcal{A}\psi^{(k)}=\lambda_k\psi^{(k)}$. Then
 the Jensen inequality with respect to $N_{11}(d\lambda)$ yields
\[\int_{-\infty}^{\infty} |{G}|_{11}^{1+\alpha/2}(x+iy)dx\le\int_{-\infty}^{\infty}dx
\int_{-\infty}^{\infty}\frac{N_{11}(d\lambda)}{(|x-\lambda|^2+y^2)^{(1+\alpha/2)/2}}
\le C|y|^{-\alpha/2}.\]
Taking $s=3/2+\alpha$ in (\ref{pj.1}) we get
\[\frac{p_n}{n}  \mathbf{ Var}\{\mathcal{N}_n[\varphi]\}\le ||\varphi||_{3/2+\alpha}^2C
\int_0^\infty
e^{-y}y^{2+2\alpha}y^{-3-\alpha}dy\le C||\varphi||_{3/2+\alpha}^2.\]
 $\square$

The next lemma is technical one. We accumulate relations which we need to prove CLT.
\begin{lemma}\label{l:*} Using notations of (\ref{repr}) we have uniformly in $z_1,z_2:\Im z_{1,2}>a$
with any $a>0$:
\begin{align}\label{*.1}
\mathbf{E}_1\{|A^\circ|^4\}&=O(p_n^{-2}),\quad \mathbf{E}_1\{|B^\circ|^4\}=O(p_n^{-2}),\\
\big(\mathbf{E}_1\{A^{-1}\}\big)^\circ&=-\Big(1+O(p_n^{-1})+O(p_n/n)\Big)
\frac{n^{-1}(\gamma_n^{(1)})^\circ}{\mathbf{E}^2\{A\}}+r,\label{*.1a}\\
\mathrm{with}\quad E\{|r^\circ|^2\}&\le  C/n^2+C/p_n^2n,\notag\end{align}
\begin{align}\label{*.1e}
p_n\mathbf{E}_1\{A^\circ(z_1)A^\circ(z_2)\}&=\frac{1}{n}\sum_{i}G^{(1)}_{ii}(z_1)G^{(1)}_{ii}(z_2)
+p_n\overset\circ{\gamma}_n^{(1)}(z_1)\overset\circ{\gamma}_n^{(1)}(z_2)/n^2, \\
p_n\mathbf{E}_1\{A^\circ(z_1)B^\circ(z_2)\}&=p_n\frac{d}{dz_2}\mathbf{E}_1\{A^\circ(z_1)A^\circ(z_2)\},\label{*.1b}\\
\mathbf{Var}\{p_n\mathbf{E}_1\{A^{\circ}(z_1)A^{\circ}(z_2)\}\}&=O(n^{-1}), \quad
\mathbf{Var}\{p_n\mathbf{E}_1\{A^{\circ}(z_1)B^{\circ}(z_2)\}\}=O(n^{-1}),\label{*.1c}\\
\mathbf{E}\{|\overset\circ\gamma_n^{(1)}(z)-\overset\circ\gamma_n(z)|^2\}&= O(p_n^{-1}).\label{*.1d}
\end{align}
Moreover,
\begin{align}\label{*.2}
&&\mathbf{Var}\{G^{(1)}_{ii}(z_1)\}= O(p_n^{-1}),\quad
|\mathbf{E}\{G^{(1)}_{ii}(z_1)\}-\mathbf{E}\{G_{ii}(z_1)\}|= O(p_n^{-1}),\\
&&|\mathbf{E}\{\gamma_n^{(1)}(z)\}/n-f(z)|= O(n^{-1}),\quad|\mathbf{E}^{-1}\{A(z)\}+f(z)|=
O(p_n^{-1}),\label{*.2a}
\end{align}
where
\begin{equation}\label{f}
    f(z)=\frac{1}{2}(\sqrt{z^2-4}-z).
\end{equation}
\end{lemma}
\textit{Proof.} Note that since $\Im z\Im (G^{(1)}m,m)\ge 0$, we can use the bound
\begin{equation}\label{b_A}|\Im A|\ge|\Im z|\Rightarrow |A^{-1}|\le|\Im z|^{-1}\le a^{-1}.\end{equation}
Relations (\ref{*.1}), (\ref{*.1e}), and (\ref{*.1b}) follow from the representation
\begin{align}\label{*.3x}
A^\circ&=\sum_{i\not=j}G^{(1)}_{ij}a_{1i}a_{1j}+\sum_{i}G^{(1)}_{ii}(a_{1i}^2)^\circ+n^{-1}
\overset\circ\gamma_n^{(1)}(z)=A_1^\circ+n^{-1}\overset\circ\gamma_n^{(1)}(z),\\
B^\circ&=\sum_{i\not=j}(G^{(1)}G^{(1)})_{ij}a_{1i}a_{1j}+\sum_{i}(G^{(1)}G^{(1)})_{ii}(a_{1i}^2)^\circ+n^{-1}
\frac{d}{dz}\overset\circ\gamma_n^{(1)}(z)\notag\\
&=B_1^\circ+n^{-1}\frac{d}{dz}\overset\circ\gamma_n^{(1)}(z), \notag\end{align} and  Lemma \ref{l:1} (see (\ref{b_var})
and (\ref{b_v.4})).

The first bound of (\ref{*.1c}) follows from (\ref{*.1e}) and (\ref{l1.1}) for $F(z)=z^2$.
The second bound of (\ref{*.1c}) follows from (\ref{*.1b}) and the first relation of (\ref{*.1c}), if we use the
fact that the variance of the derivative of an analytic function by the Cauchy theorem can be bounded
by the variance of the initial function.

Relations (\ref{*.1d}) follow from the representation (see (\ref{rep_1}))
\[\overset\circ\gamma_n^{(1)}(z)-\overset\circ\gamma_n(z)=(A^{-1})^\circ+(BA^{-1})^\circ\]
and (\ref{*.3}).
The first relation of (\ref{*.2}) is the
analog of the relation
\begin{equation}\label{*.3a}
\mathbf{Var}\{G_{ii}(z_1)\}=\mathbf{Var}\{G_{11}(z_1)\}= O(p_n^{-1})\end{equation}
if in the latter we replace the matrix $\mathcal{A}$ by $\mathcal{A}^{(1)}$. But since $G_{11}(z_1)=-A^{-1}(z_1)$,
(\ref{*.3a}) follows from (\ref{*.1}) and  (\ref{b_A}).  The second relation of (\ref{*.2})
follows from the symmetry of the problem and (\ref{repr})
\begin{align*}
\mathbf{E}\{G^{(1)}_{ii}(z_1)\}-\mathbf{E}\{G_{ii}(z_1)\}=\frac{1}{n-1}
\mathbf{E}\{\gamma_n-\gamma_n^{(1)}-G_{11}\}=\frac{1}{n-1}\mathbf{E}\{B/A\}=O(n^{-1}).
\end{align*}
The first relation of (\ref{*.2a}) follows from the above bound for $n^{-1}\mathbf{E}\{\gamma_n-\gamma_n^{(1)}\}$
and the  estimate (see \cite{KKPS})
\[n^{-1}\mathbf{E}\{\gamma_n\}-f(z)=O(p_n^{-1}).\]
The second relation of (\ref{*.2a}) is the corollary of the above estimate and the  representation
\begin{equation}\label{*.3}
    A^{-1}=\mathbf{E}^{-1}\{A\}-A^\circ\mathbf{E}^{-2}\{A\}+
(A^\circ)^2A^{-1}\mathbf{E}^{-2}\{A\},
\end{equation}
which implies
\[\mathbf{E}\{A(z)\}^{-1}=\mathbf{E}\{A(z)^{-1}\}+O(\mathbf{Var}\{A(z)\})=-\mathbf{E}\{G_{11}\}+O(p_n^{-1})=
-n^{-1}\mathbf{E}\{\gamma_n\}+O(p_n^{-1}).\]
We are left to prove (\ref{*.1a}). Set
\begin{equation}\label{ti_A}
\widetilde A=z+\sum G^{(1)}_{ii}a_{1i}^2.
\end{equation}
  Using the analog of (\ref{*.3}) for $A$ and $\widetilde A$, we write first
\[A^{-1}=\widetilde A^{-1}-\widetilde A^{-2}(A-\widetilde A)+r_1,
\quad r_1=\widetilde A^{-2}A^{-1}(A-\widetilde A)^2.
\]
We have
\[A-\widetilde A=\sum_{i\not=j}G^{(1)}_{ij}a_{1i}a_{1j},\quad \mathbf{E}\{|r_1|^2\}\le\frac{1}{|\Im z|^6}
\mathbf{E}\{|A-\widetilde A|^4\}\le C/n^2+C/np_n^2.\]
Moreover, the analog of (\ref{*.3}) for $\widetilde A$ yields
\begin{align*}\notag
\mathbf{E}_1\{\widetilde A^{-2}(A-\widetilde A)\}&=\mathbf{E}_1^{-2}\{A\}\mathbf{E}_1\{(A-\widetilde A)\}
\\&-2\mathbf{E}_1^{-3}\{A\}\mathbf{E}_1\{(A-\widetilde A)(\widetilde A-\mathbf{E}_1\{A\})\}+r_2=r_2\\
r_2&=\mathbf{E}_1\Big\{\mathbf{E}_1^{-2}\{A\}\Big(\widetilde A^{-2}+2\mathbf{E}_1^{-1}\{A\}\widetilde A^{-1}
\Big)\Big(A-\widetilde A\Big)\Big(\widetilde A-\mathbf{E}_1\{A\}\Big)^2\Big\}.\end{align*} Since $\widetilde
A-\mathbf{E}_1\{A\}=\sum G_{ii}^{(1)}(a_{1i}^2-\mathbf{E}_1\{a_{1i}^2\})$ we have
\begin{equation*}
    \mathbf{E}\{r_2^2\}\le C\mathbf{E}\{|A-\widetilde A|^2|\widetilde
A-\mathbf{E}_1\{A\}|^4\}\le C/np_n^2.
\end{equation*}
Hence we have proved that
\begin{equation}\label{*.4}
\mathbf{E}_1\{ A^{-1}\}=\mathbf{E}_1\{\widetilde A^{-1}\}+\widetilde r,
\quad E\{|\widetilde r|^2\}\le  C/n^2+C/np_n^2.
\end{equation}
Using (\ref{A}) we can write
\begin{align*}
i\mathbf{E}_1\{\widetilde A^{-1}\}&=\int_0^\infty dte^{izt}\prod\mathbf{E}_1\{ e^{itG^{(1)}_{jj}a_{1j}^2}\}
\\&=\int_0^\infty
dte^{izt}e^{itp_n\gamma_n^{(1)}/n^2}\prod\Big(1+\frac{p_n}{n}\Big(e^{itG^{(1)}_{ii}(1/p_n-2/n)}-1\Big)\Big)
\\&=\int_0^\infty dte^{izt}e^{itp_n\gamma_n^{(1)}/n^2}\exp\Big\{\frac{p_n}{n}\sum_{i}
\Big(e^{itG^{(1)}_{ii}(1/p_n-2/n)}-1\Big)
\Big\}+O(n^{-1})
\\&=\int_0^\infty dte^{izt} e^{it\gamma_n^{(1)}(1-p_n/n)/n}\exp\Big\{in^{-1}\sum
F(G^{(1)}_{ii},t)\Big\}+O(n^{-1}),
\end{align*}
where
\[  F(x,t)= p_n\Big(e^{itx(1/p_n-2/n)}-1-itx(1/p_n-2/n) \Big) , \]
Then in view of (\ref{l1.1}), since $\sup_{\Im x>0}|F'_x(x,t)|\le C|t|p_n^{-1}$, we obtain
\begin{align*}
i\mathbf{E}_1\{\widetilde A^{-1}\}&=\int_0^\infty dte^{izt} e^{it\gamma_n^{(1)}(1-p_n/n)/n}\exp\Big\{in^{-1}\sum
\mathbf{E}\{F(G^{(1)}_{ii},t)\}\Big\}\\&\cdot\Big(1+O\Big(n^{-1}\sum
F^\circ(G^{(1)}_{ii},t)\Big)\Big)+O(n^{-1})\\
&=\int_0^\infty dte^{izt}e^{it\gamma_n^{(1)}(1-p_n/n)/n}\exp\Big\{in^{-1}\sum
\mathbf{E}\{F(G^{(1)}_{ii},t)\}\Big\}+r',\\
&\mathbf{E}\{|r'|^2\}\le C/n^2+ C/np_n^2.
\end{align*}
Finally, replacing similarly to the above $\gamma_n^{(1)}$ by $\mathbf{E}\{\gamma_n^{(1)}\}$ in the exponent, we get
\begin{align*}
i\mathbf{E}_1\{\widetilde A^{-1}\}&=\int_0^\infty dte^{izt}e^{it\mathbf{E}\{\gamma_n^{(1)}\}(1-p_n/n)/n}
\exp\Big\{in^{-1}\sum
\mathbf{E}\{F(G^{(1)}_{ii},t)\}\Big\}\\
&\Big(1+itn^{-1}(\gamma_n^{(1)})^\circ(1-p_n/n)+O((n^{-1}(\gamma_n^{(1)})^\circ)^2\Big)+r'.
\end{align*}
Taking $(\mathbf{E}_1\{\widetilde A^{-1}\})^\circ $, we can see that
the term which corresponds to 1 in the r.h.s.  disappears, and since
$\mathbf{E}\{F(G_{ii},t)\}=O(p_n^{-1})$,
the coefficient in front of $(\gamma_n^{(1)})^\circ$ equals
\begin{align*}
&&\int_0^\infty ite^{izt}e^{it\mathbf{E}\{\gamma_n^{(1)}\}(1-p_n/n)/n}
\exp\Big\{in^{-1}\sum
\mathbf{E}\{F(G^{(1)}_{ii},t)\}\Big\} dt\\
&&=i(z+\mathbf{E}\{\gamma_n^{(1)}\}/n)^{-2}(1+O(p_n^{-1})+O(p_n/n))\\&&=-i\mathbf{E}^{-2}\{A\}
\big(1+O(p_n^{-1})+O(p_n/n)\big).
\end{align*}
In view of (\ref{b_var}) and (\ref{*.4}) we obtain (\ref{*.1a}).

$\square$

\textit{Proof of Theorem \ref{t:CLT}}. We prove first Theorem \ref{t:CLT} for the  function
$\varphi_\eta$ of the form
\begin{equation}\label{phi_e}
\varphi_\eta=P_\eta*\varphi_0,\quad \int|\varphi_0(\lambda)|d\lambda\le C<\infty,
\end{equation}
where $P_\eta$ is the Poisson kernel (see (\ref{P_y})) and $\varphi_0$ is a real valued function
from $L_1(\mathbb{R})$.
 One can see easily that
\begin{align}\notag
\mathcal{N}_n^\circ[\varphi_\eta]&= \Big(\mathrm{Tr\,
}\varphi_\eta(\mathcal{A})\Big)^{\circ}=\frac{1}{\pi}\int\varphi_0(\mu)
\Im\Big(\mathrm{Tr\, }G(\mu+i\eta)\Big)^{\circ}d\mu\\
&= \frac{1}{2\pi i}\int \varphi_0(\mu)(\gamma_n^\circ(z_\mu)-\gamma_n^\circ(\overline z_\mu))d\mu,\quad
z_\mu=\mu+i\eta. \label{repr_N}\end{align} Set
\begin{align}\label{Z_n}
d_n&=\Big(\frac{n}{p_n}\Big)^{1/2},\quad Z_n(x)=\mathbf{E}\{e^{ix\mathcal{N}_n^\circ[\varphi_\eta]/d_n}\},\quad\\
e(x)&=e^{ix\mathcal{N}_n^\circ[\varphi_\eta]/d_n}, \quad Y_n(z,x)=d_n^{-1}\mathbf{E}\{\hbox{Tr}G(z)e^\circ(x)\}.
\notag\end{align} Then it is easy to see that
\begin{align}\label{CLT.1}
\frac{d}{dx}Z_n(x)&=\frac{1}{2\pi}\int \varphi_0(\mu) (Y_n(z_\mu,x)-Y_n(\overline z_\mu,x))d\mu.
\end{align}
On the other hand, using the symmetry of the problem and
the notations of (\ref{repr}), we have
\begin{align}\label{Y=}
Y_n(z,x)&=d_n^{-1}\mathbf{E}\{\mathrm{Tr\,} G(z)e^\circ(x)\}=nd_n^{-1}\mathbf{E}\{G_{11}(z)e^\circ(x)\}\\
&= -nd_n^{-1}\mathbf{E}\{(A^{-1})^\circ e_1(x)\}-nd_n^{-1}\mathbf{E}\{(A^{-1})^\circ (e(x)-e_1(x))\}=T_1+T_2,
\notag\end{align} where
\[e_1(x)=e^{ix(\mathcal{N}_{n-1}^{(1)}[\varphi_\eta])^\circ/d_n}, \quad
(\mathcal{N}_{n-1}^{(1)}[\varphi_\eta])^\circ=(\hbox{Tr}\varphi_\eta(\mathcal{A}^{(1)}))^\circ
=\frac{1}{\pi}\int d\mu\,\varphi_0(\mu)\Im\overset\circ\gamma_n^{(1)}(z_\mu).\]
Since $e_1(x)$ does not depend on $\{a_{1i}\}$, using that $\mathbf{E}\{...\}=\mathbf{E}\{\mathbf{E}_1\{...\}\}$, we
obtain in view of the above representation  and  (\ref{*.1a})
\[T_1=d_n^{-1}\mathbf{E}\{(\gamma^{(1)}_n(z))^\circ e_1(x)\}/\mathbf{E}^2\{A\}\Big(1+O(d_n^{-1})
+O(p_n^{-1})\Big)+O(d_n^{-1}).\]
Write
\begin{align}\label{e-e_1}
    e(x)-e_1(x)=\frac{ix}{\pi d_n}\int\varphi_0(\mu)\Big(\Im\Big(\gamma_n^\circ-\overset\circ{\gamma_n}^{(1)}\Big)
    e_1(x)
+O\Big((\gamma_n^\circ-\overset\circ{\gamma_n}^{(1)})^2\Big)ix/d_n\Big)d\mu.
\end{align}
Then  (\ref{*.1d}),  the relations $|e(x)|=|e_1(x)|=1$, and (\ref{b_var}) yield
\begin{align*}
d_n^{-1}|\mathbf{E}\{(\gamma^{(1)}_n)^\circ e_1(x)\}-\mathbf{E}\{\gamma_n^\circ e(x)\}|\le
Cd_n^{-1}\mathbf{E}\{|(\gamma^{(1)}_n)^\circ-\gamma_n^\circ|(1+|x||\gamma_n^\circ|d_n^{-1})\}\\
\le Cd_n^{-1}\mathbf{E}^{1/2}\{|(\gamma^{(1)}_n)^\circ-\gamma_n^\circ|^2\}
\Big(1+|x|d_n^{-1}\mathbf{E}^{1/2}\{|\gamma_n^\circ|^2\}\Big)=O(d_n^{-1}p_n^{-1/2}).\notag
\end{align*}
Hence we obtain
\begin{equation}\label{T_1}
T_1=Y_n(z,x)/\mathbf{E}^2\{A\}\Big(1+O(d_n^{-1})+O(p_n^{-1})\Big)+O(d_n^{-1}p_n^{-1/2}).
\end{equation}
To compute $T_2$  we use (\ref{e-e_1}). Then, taking into account
 (\ref{*.1d}),  we conclude that the term
$O(nd_n^{-3}(\gamma_n^\circ-(\gamma^{(1)}_n)^\circ)^2)$ gives the contribution $O(d_n^{-1})$. Then, since $e_1(x)$
does not depend on $\{a_{1i}\}$, we  average first with respect to $\{a_{1i}\}$ and obtain
in view of (\ref{repr})
\begin{align*}
T_2&=-\frac{ix n}{d_n^2\pi}\int d\mu\varphi_0(\mu)\mathbf{E}\bigg\{(A^{-1})^\circ(z)
e_1(x)\Im\Big(\gamma_n^\circ(z_\mu)-(\gamma^{(1)}_n(z_\mu))^\circ\Big)\bigg\}+O(d_n^{-1})\\
&=-\frac{ixp_n}{\pi}\int d\mu\varphi_0(\mu)\mathbf{E}\bigg\{e_1(x)\mathbf{E}_1\Big\{(A^{-1})^\circ(z)
\Im\Big(\gamma_n^\circ(z_\mu)-(\gamma^{(1)}_n(z_\mu))^\circ\Big)\Big\}\bigg\}+O(d_n^{-1})\\
&=\frac{ix p_n}{\pi}\int d\mu\varphi_0(\mu)\mathbf{E}\bigg\{e_1(x)\mathbf{E}_1\Big\{(A^{-1})^\circ(z)
\Im\Big((1+B(z_\mu))A^{-1}(z_\mu)\Big)^\circ\Big\}\bigg\}+O(d_n^{-1}).
\end{align*}
Using (\ref{*.3}) and (\ref{*.1}), we conclude that only linear terms with respect to $B^\circ$
and $A^\circ$  give non vanishing contribution, hence
we obtain
\begin{align*}
D_n(z,z_\mu)&:=p_n\mathbf{E}_1\Big\{(A^{-1})^\circ(z)\Big((1+B(z_\mu))A^{-1}(z_\mu)\Big)^\circ\Big\}\\
&=p_n\mathbf{E}^{-2}\{A(z)\}\mathbf{E}^{-2}\{A(z_\mu)\}
\Big(1+\mathbf{E}\{B(z_\mu)\}\Big)\mathbf{E}_1\{A^\circ(z)A^\circ(z_\mu)\}\\
&-p_n\mathbf{E}^{-2}\{A(z)\}\mathbf{E}^{-1}\{A(z_\mu)\}
\mathbf{E}_1\{A^\circ(z)B^\circ(z_\mu)\}+O(p_n^{-1/2})\\
&=f^3(z)f^3(z_\mu)(1+f'(z_\mu))+f^3(z)f(z_\mu)f'(z_\mu)+O(p_n^{-1/2}).
\end{align*}
Here we used first (\ref{*.1e}) and (\ref{*.1b}) to express  $\mathbf{E}_1\{A^\circ(z)A^\circ(z_\mu)\}$
and $\mathbf{E}_1\{A^\circ(z)B^\circ(z_\mu)\}$ in terms of $G^{(1)}_{ii}(z)$, and
$\frac{d}{dz_\mu}G^{(1)}_{ii}(z_\mu)$, and then \ref{*.2})
combined with (\ref{*.2a}) to replace $\mathbf{E}_1\{A^\circ(z)A^\circ(z_\mu)\}$ by $f(z)f(z_\mu)$ and
$\mathbf{E}_1\{A^\circ(z)B^\circ(z_\mu)\}$ by $f(z)f'(z_\mu)$. Moreover,
we used (\ref{*.2a}) to replace $\mathbf{E}^{-1}\{A(z)\}$ by $-f(z)$.
 Hence
\begin{align}\label{D_n}
D_n(z,z_\mu)&=\Big(f^3(z)f^3(z_\mu)(1+f'(z_\mu))+f^3(z)f(z_\mu)f'(z_\mu)\Big)+O(p_n^{-1/2}).
\end{align}
In addition, similarly to (\ref{e-e_1}) we have
\[\mathbf{E}\{e_1(x)\}=Z_n(x)+O(d_n^{-1}).\]
Hence, relations (\ref{Y=})--(\ref{D_n}) imply
\begin{align}\notag
Y_n(z,x)&=f^2(z)Y_n(z,x)+ixZ_n(x)\int d\mu\varphi_0
(\mu)\frac{D_n(z,z_\mu)-D_n(z,\overline{z_\mu})}{2i\pi}\\
&+O(p_n^{-1/2}) +O(d_n^{-1}),\notag\\\label{Y} Y_n(z,x)&=ixZ_n(x)\int d\mu\varphi_0
(\mu)\frac{C_n(z,z_\mu)-C_n(z,\overline{z_\mu})}{2i\pi} +O(p_n^{-1/2})
+O(d_n^{-1}),\\
 C_n(z,z_\mu):&=\frac{D_n(z,z_\mu)}{1-f^2(z)}.
\notag\end{align} Using that
\[
f(z)(f'(z)+1)=\frac{f(z)}{1-f^2(z)}=-\frac{1}{\sqrt{z^2-4}},\quad
f'=-\frac{f(z)}{\sqrt{z^2-4}},\]
we can transform $C_n(z,z_\mu)$ to the form
\begin{align}\label{C}
C_n(z,z_\mu)&=C(z,z_\mu)+O(p_n^{-1/2})+O(d_n^{-1/2})\\
C(z,z_\mu):&=2\frac{f^2(z)f^2(z_\mu)}{(z^2-4)^{1/2}(z_\mu^2-4)^{1/2}}.\notag\end{align}
Taking into account
(\ref{CLT.1}), (\ref{Y}), and (\ref{C}), we obtain the equation
\begin{align}\label{equ}
\frac{d}{dx}Z_n(x)&=-xV[\varphi_\eta]Z_n(x)+O(p_n^{-1/2})+O(d_n^{-1})\\
V[\varphi_\eta]&=-\frac{1}{4\pi^2}\int\int\varphi_0(\mu_1)\varphi_0(\mu_2)\Big(
C(z_{\mu_1},z_{\mu_2})-C(z_{\mu_1},\overline{z_{\mu_2}})-C(\overline{z_{\mu_1}},z_{\mu_2})\notag\\&+
C(\overline{z_{\mu_1}},\overline{z_{\mu_2}})\Big)d\mu_1 d\mu_2. \notag\end{align}
Formulas (\ref{C}) and (\ref{equ})
imply that
\begin{align*}
V[\varphi_\eta]=\frac{2}{\pi^2}\bigg(\int\varphi_0(\mu)\Im\frac{f^2(z_\mu)}{(z^2_\mu-4)^{1/2}}d\mu\bigg)^2=
\frac{1}{2\pi^2}\bigg(\int\varphi_0(\mu)\Im\Big(\frac{z_\mu^2-2}{(z^2_\mu-4)^{1/2}}-z_\mu\Big)d\mu\bigg)^2\\
=\frac{1}{2\pi^2}\bigg(\int d\mu\varphi_0(\mu)
\Im\Big(\frac{1}{\pi}\int_{-2}^2\frac{(\lambda^2-2)d\lambda}{(\mu+i\eta-\lambda)\sqrt{4-\lambda^2}}\Big)\bigg)^2
=\frac{1}{2\pi^2}\bigg(\int_{-2}^2 d\lambda\varphi_\eta(\lambda)
\frac{(\lambda^2-2)}{\sqrt{4-\lambda^2}}\bigg)^2,
\end{align*}
where we  used also the well known relations
\[\frac{1}{\pi}\int_{-2}^2\frac{d\lambda}{(z-\lambda)\sqrt{4-\lambda^2}}=\frac{1}{(z^2-4)^{1/2}},
\quad \frac{1}{\pi}\int_{-2}^2\frac{d\lambda}{\sqrt{4-\lambda^2}}=1.\]

Now if we consider
\[\widetilde Z_n(x)=e^{x^2V[\varphi_\eta]/2}Z_n(x),\]
then (\ref{equ}) yields that for any $|x|\le C$
\[\frac{d}{dx}\widetilde Z_n(x)=O(p_n^{-1/2})+O(d_n^{-1}),\]
and since $\widetilde Z_n(0)=Z_n(0)=1$, we obtain uniformly in $x\le C$
\begin{align}\notag
&\widetilde Z_n(x)=1+O(p_n^{-1/2})+O(d_n^{-1})\\
 \Rightarrow
 &Z_n(x)=e^{-x^2V[\varphi_\eta]/2}+O(p_n^{-1/2})+O(d_n^{-1}).
\label{conv_Z}\end{align} Thus, we have proved CLT for the functions of the form (\ref{phi_e}). To extend CLT to a
wider class of functions we use
\begin{proposition} \label{p:CLTcont}
Let $\{\xi_{l}^{(n)}\}_{l=1}^{n}$ be a triangular array of random variables,
$\displaystyle\mathcal{N}_{n}[\varphi ]=\sum_{l=1}^{n}\varphi
(\xi _{l}^{(n)})$ be its linear statistics,
corresponding to a test function $\varphi :\mathbb{R}\rightarrow
\mathbb{R}$, and
\[V_{n}[\varphi]=\mathbf{Var}\{d_n^{-1}\mathcal{N}_{n}[\varphi ]\}\]
be the variance of $\mathcal{N}_{n}[\varphi ]$, where $\{d_n\}_{n=1}^\infty$ is some bounded from above sequence
of numbers. Assume that

(a) there exists a vector space $\mathcal{L}$ endowed with a norm $||...||$
and such that $V_{n}$ is defined on $\mathcal{L}$ and admits the bound
\begin{equation}
V_{n}[\varphi ]\leq C||\varphi ||^2,\quad \forall \varphi \in \mathcal{L}, \label{ocVn1}
\end{equation}%
where $C$ does not depend on $n$;

(b) there exists a dense linear manifold $\mathcal{L}_{1}\subset \mathcal{L}$ such that the Central Limit Theorem is
valid for $\mathcal{N}_{n}[\varphi ],\quad \varphi \in \mathcal{L}_{1}$, i.e., if
$Z_{n}[x\varphi ]=\mathbf{E}\big\{
e^{ixd_n^{-1}\overset{\circ}{\mathcal{N}}_{n}[\varphi ]}\big\}$
is the characteristic function of
$d_n^{-1/2}\overset{\circ }{\mathcal{N}}_{n}[\varphi ]$, then there
exists a continuous quadratic functional $V:\mathcal{L}_{1}\rightarrow \mathbb{R}_{+}$
such that we have uniformly in $x$,
varying on any compact interval
\begin{equation}\label{limZ}
\lim_{n\rightarrow \infty }Z_{n}[x\varphi ]=e^{-x^{2}V[\varphi ]/2},\quad \forall \varphi \in \mathcal{L}_{1};
\end{equation}
Then  $V$ admits a continuous extension to $\mathcal{L}$ and Central Limit
Theorem is valid for all $\mathcal{N}_{n}[\varphi ]$,
 $\varphi \in \mathcal{L}$.
\end{proposition}

\textit{Proof.} Let $\{\varphi _{k}\}$ be a sequence of elements of
$\mathcal{L}_{1}$ converging to $\varphi \in \mathcal{L}$. We have
then in view of the inequality $ |e^{ia}-e^{ib}|\leq |a-b|$, the
linearity of $\overset{\circ}{\mathcal{N}} _{n}[\varphi ]$ in
$\varphi $, the Schwarz inequality, and (\ref{ocVn1}):
\begin{align} \label{difZ}\Big|Z_{n}(x\varphi
)-Z_{n}(x\varphi)|_{\varphi=\varphi_k} \Big| &\leq |x|\mathbf{E}\left\{ \left\vert
d_n^{-1}\overset{\circ}{\mathcal{N}}_{n}[\varphi ]-d_n^{-1}\overset{\circ}{ \mathcal{N}}_{n}[\varphi _{k}]\right\vert
\right\} \\ &\leq |x|\mathbf{Var}^{1/2}\{d_n^{-1}\mathcal{N}_{n}[\varphi -\varphi _{k}]\}\leq C|x|\quad ||\varphi
-\varphi _{k}||.\notag
\end{align}
Now,
passing first to the limit $n\rightarrow \infty $ and then
$k\rightarrow \infty $, we obtain the assertion of the proposition. $\square$

\medskip

Let us show now that hypothesis (a) and (b) of Proposition \ref{p:CLTcont} are fulfilled
in some vector space. Consider the space $\mathcal{H}_s$ of all functions with the norm
(\ref{norm}) and set $\mathcal{L}=\mathcal{H}_s\cap L_1(\mathbb{R})$ and
\begin{equation}\label{norm1}
||\varphi||=\int|\varphi(\lambda)|d\lambda+||\varphi||_s=||\varphi||_{L_1(\mathbb{R})}+||\varphi||_s.
\end{equation}
 Then for  $s>3/2$ Lemma \ref{l:2} guarantees that assumption (a) of Proposition \ref{p:CLTcont}
 is fulfilled. Moreover, the Lebesgue theorem about the dominated convergence yields that
 \[||\varphi-\varphi* P_\eta||^2_s\le C\int|1-e^{-\eta|k|}|^2(1+2|k|)^{2s}|
 \widehat\varphi(k)|^2dk\to0,\quad\eta\to0.\]
 Hence the set of the functions $\varphi* P_\eta$ is dense in $\mathcal{L}$ with respect to the norm
 $||.||_s$. Thus, if we prove that the set of the functions $\varphi* P_\eta$
 is dense in $\mathcal{L}$ with respect to the norm $||.||_{L_1(\mathbb{R})}$, then (\ref{conv_Z}) will imply
assumption (b) of Proposition \ref{p:CLTcont}.

It is easy to see that the set of all functions with finite supports, possessing the norm
(\ref{norm1}), is dense in $\mathcal{L}$ with respect to this norm. Hence we need only to prove
that if $\varphi\in\mathcal{H}_s$ and has a finite support $[-A,A]$, then
\[\int|\varphi(\lambda)-\varphi* P_\eta(\lambda)|d\lambda\to 0,\quad\eta\to 0.\]
But
\[\int|\varphi(\lambda)-\varphi* P_\eta(\lambda)|d\lambda=\bigg(\int_{|\lambda|\le A+1}
+\int_{|\lambda|\ge A+1}\bigg)|\varphi(\lambda)-\varphi*
P_\eta(\lambda)|d\lambda=I_1+I_2.\]
We have for $I_2$
\begin{align}\label{b_I2}
I_2\le
\frac{\eta}{\pi}\int_{|\lambda|\ge A+1}d\mu
\int_{|\lambda|\le A}\frac{|\varphi(\lambda)|d\lambda}{(\lambda-\mu)^2+\eta^2}\le C\eta||\varphi||_{L_1(\mathbb{R})},
\end{align}
and for $I_1$ we use the inequalities:
\begin{equation}\label{b_I1}
I_1\le 2|A+1|\sup_{|\lambda|\le A+1}|\varphi(\lambda)-\varphi* P_\eta(\lambda)|\le C|A+1|
\int|1-e^{-\eta|k|}||\widehat\varphi(k)|dk.
\end{equation}
But since
\[\int|\widehat\varphi(k)|dk=\int\frac{|\widehat\varphi(k)|(1+2|k|)^s}{(1+2|k|)^s}dk\le
||\varphi||_s\Big(\int\frac{dk}{(1+2|k|)^{2s}}\Big)^{1/2}\le C||\varphi||_s,\]
(\ref{b_I1})  the Lebesgue theorem on the dominated convergence implies that $I_1\to 0$, as
$\eta\to 0$. Combining this with (\ref{b_I1}) we get that the set of all functions with
finite supports, possessing the norm
(\ref{norm1}) is dense in $\mathcal{L}$.  As it was mentioned above this implies that
the set of the functions $\varphi* P_\eta$ with $\varphi\in\mathcal{L}$
 is dense in $\mathcal{L}$ with respect to the norm $||.||_{L_1(\mathbb{R})}$ and in view of (\ref{conv_Z})
 proves assumption (b) of the proposition.
$\square$

\medskip
\textit{Proof of Theorem \ref{t:CLT2}.} Let us note first that  in the case when
\begin{equation}\label{ineq_p}
    p_n>C_*\log^{1/3} n,\quad n\to\infty,
\end{equation}
the proof of  Theorem \ref{t:CLT2} is rather simple.  By the method of \cite{BdMS}  one can prove the estimate
\[ n^{-1}\mathbf{E}\{\mathrm{Tr\,}  \mathcal{A}^{2m}\}\le C^2(1+\frac{m^3}{p_n})
n^{-1}\mathbf{E}\{\mathrm{Tr\,}  \mathcal{A}^{2m-2}\}.\]
Then under condition (\ref{ineq_p})  it is easy to get the
bound, valid for sufficiently big $K$:
\[\mathbf{Prob}\{|| \mathcal{A}||\ge K\}\le \inf_m n\mathbf{E}\{\mathrm{Tr\,}  (\mathcal{A}/K)^{2m}\}\le
\exp\{-p_n^{1/3}\log(K/2C)+\log n\}\to 0, n\to\infty.\] Then, for any $\varphi\in\mathcal{H}^{(c)}_s$, if we consider a
smooth  function $\varphi^{(K)}\in \mathcal{H}_s$ with a finite support and such that
$\varphi^{(K)}(\lambda)=\varphi(\lambda)$, $|\lambda|\le K$, then evidently
\begin{align*}|\mathbf{E}\{e^{ix\mathcal{N}_n^\circ[\varphi]/d_n}\}&-
\mathbf{E}\{e^{ix\mathcal{N}_n^\circ[\varphi^{(K)}]/d_n}\}|\le \mathbf{Prob}\{|| \mathcal{A}||
K\}\\&+d_n^{-1}|\mathbf{E}\{\mathcal{N}_n[\varphi]\}-\mathbf{E}\{\mathcal{N}_n[\varphi^{(K)}]\}|\to 0, \quad
n\to\infty.\end{align*}
 Thus, we can
derive Theorem \ref{t:CLT2} from Theorem \ref{t:CLT} almost immediately.

\smallskip

But if the inequality (\ref{ineq_p}) is not fulfilled, then the proof of Theorem \ref{t:CLT2} is more complicated. It
is based on the bound which is the analog of (\ref{pj.1})
\begin{equation}\label{main_1}
   \frac{p_n}{n} \mathbf{Var}\{\mathcal{N}_n[\varphi]\}\le C(c)||\widetilde \varphi||_{s},\quad
\varphi\in\mathcal{H}_s^{(c)},\end{equation}
where $ \widetilde\varphi(\lambda)=\varphi(\lambda)\cosh^{-1}(c\lambda)$.
The main step here is  the lemma, which is the generalization of Lemma \ref{l:1}
\begin{lemma}\label{l:4} Denote by $\gamma_n^{(c)}=\mathrm{Tr\,} G(z)e^{c\mathcal{A}}$. Then for any $1>\varepsilon>0$
\begin{equation}\label{l4.1}
    \frac{p_n}{n}\mathbf{Var}\{\gamma_n^{(c)}\}\le
    C(c,\varepsilon)\mathbf{E}\{|G_{11}|^{1+\varepsilon}\}/|\Im z|^{3+\varepsilon}.
\end{equation}
\end{lemma}
\textit{Proof.} According to Proposition \ref{p:mart} it is enough to prove that
\begin{equation}\label{l4.1a}
\mathbf{E}\{|\gamma_n^{(c)}-\mathbf{E}_1\{\gamma_n^{(c)}\}|^2\}\le
C(c,\varepsilon)\mathbf{E}\{|G_{11}|^{1+\varepsilon}\}/|\Im z|^{3+\varepsilon}p_n.
\end{equation}
Let us  set
\[G^{(1)}(z)=(\mathcal{A}^{(1)}-z)^{-1},\quad\gamma_n^{(1c)}=\mathrm{Tr\,} G^{(1)}(z)e^{c\mathcal{A}^{(1)}}.\]
 Note that differently from the proof of
Proposition \ref{p:mart} here and below we denote by $\mathcal{A}^{(1)}$ the $n\times n$ matrix whose first line and
column are zero and the other entries coincide with those of $\mathcal{A}$. We also  denote
$a^{(1)}=(0,a_{12},\dots,a_{1n})$.
Then we can write
\begin{align*}
\gamma_n^{(c)}-\mathbf{E}_1\{\gamma_n^{(c)}\}=&\gamma_n^{(c)}-\gamma_n^{(1c)}-\mathbf{E}_1\{\gamma_n^{(c)}-\gamma_n^{(1c)}\},\\
\gamma_n^{(c)}-\gamma_n^{(1c)}=&\mathrm{Tr\,} (G(z)- G^{(1)}(z))e^{c\mathcal{A}^{(1)}}+\mathrm{Tr\,}
G^{(1)}(z)(e^{c\mathcal{A}}-e^{c\mathcal{A}^{(1)}})\\&+ \mathrm{Tr\,} (G(z)-
G^{(1)}(z))(e^{c\mathcal{A}}-e^{c\mathcal{A}^{(1)}})=I+II+III.
\end{align*}
Let us use the formulas
\begin{align*}
(G(z)- G^{(1)}(z))_{11}&=z^{-1}-A^{-1},\quad (G(z)- G^{(1)}(z))_{1i}=-A^{-1}(G^{(1)}a^{(1)})_i,\\
(G(z)- G^{(1)}(z))_{ij}&=-A^{-1}(G^{(1)}a^{(1)})_i(G^{(1)}a^{(1)})_j,\quad i,j\ge 2,\\
(e^{c\mathcal{A}}-e^{c\mathcal{A}^{(1)}})_{11}&=(e^{c\mathcal{A}})_{11}-1,\quad
(e^{c\mathcal{A}}-e^{c\mathcal{A}^{(1)}})_{1i}=
c\int_0^1dt(e^{c(1-t)\mathcal{A}})_{11}(e^{ct\mathcal{A}^{(1)}}a^{(1)})_i,\\
 (e^{c\mathcal{A}}-e^{c\mathcal{A}^{(1)}})_{ij}&=c^2\int_0^1dt\int_0^{1-t}d\tau
(e^{ct\mathcal{A}^{(1)}}a^{(1)})_i (e^{c\tau\mathcal{A}^{(1)}}a^{(1)})_j(e^{c(1-t-\tau)\mathcal{A}})_{11},\quad i,j\ge
2.
\end{align*}
Here $A$ is defined in (\ref{repr}) and  to obtain the last two lines  we have used the Duhamel formula, valid for any
matrices $\mathcal{M}$ and $\mathcal{M}^{(1)}$:
\[e^{c\mathcal{M}}-e^{c\mathcal{M}^{(1)}}=c\int_0^1dte^{c\mathcal{M}^{(1)}t}(\mathcal{M}-\mathcal{M}^{(1)})e^{c\mathcal{M}(1-t)}.\]
Moreover, we have taken into account that
\[\mathcal{A}^{(1)}_{11}=\mathcal{A}^{(1)}_{i1}=G^{(1)}_{i1}=(e^{t\mathcal{A}^{(1)}})_{i1}=0,\quad \mathcal{A}^{(1)}_{11}=0,
\quad G^{(1)}_{11}=-z^{-1},\quad (e^{t\mathcal{A}^{(1)}})_{11}=1.\] Hence we have
\begin{align*}
I=&z^{-1}-A^{-1}-A^{-1}(e^{c\mathcal{A}^{(1)}}(G^{(1)})^2a^{(1)},a^{(1)}),\\
 II=&-z^{-1}((e^{c\mathcal{A}})_{11}-1)+c^2\int_0^1dt\int_0^{1-t}d\tau(e^{c(1-t-\tau)\mathcal{A}})_{11}
(G^{(1)}e^{ct\mathcal{A}^{(1)}}a^{(1)},e^{c\tau\mathcal{A}^{(1)}}a^{(1)})\\
=&-z^{-1}((e^{c\mathcal{A}})_{11}-1)+c^2\int_0^1s(e^{cs\mathcal{A}^{(1)}}G^{(1)}a^{(1)},a^{(1)})
(e^{c(1-s)\mathcal{A}})_{11}ds, \\
III=& ((e^{c\mathcal{A}})_{11}-1)(z^{-1}-A^{-1})-2cA^{-1}\int_0^1dt
(e^{ct\mathcal{A}^{(1)}}G^{(1)}a^{(1)},a^{(1)})(e^{c(1-t)\mathcal{A}})_{11}\\&-c^2A^{-1}\int_0^1dt\int_0^{1-t}d\tau
(e^{ct\mathcal{A}^{(1)}}G^{(1)}a^{(1)},a^{(1)})
(e^{c\tau\mathcal{A}^{(1)}}G^{(1)}a^{(1)},a^{(1)})(e^{c(1-t-\tau)\mathcal{A}})_{11}.
\end{align*} Thus, denoting
$B^{(c)}:=(e^{c\mathcal{A}^{(1)}}G^{(1)}a^{(1)},G^{(1)}a^{(1)})$ and using the Schwarz inequality and
(\ref{in_a})-(\ref{in_a1}), we get for $I$
\begin{align*}
\mathbf{E}\{|I-\mathbf{E}_1\{I\}|^2\}\le 3\mathbf{E}\{|A^{-1}-\mathbf{E}_1\{A^{-1}\}|^2\}+
3\mathbf{E}\bigg\{\bigg|\frac{B^{(c)\circ}_1}{\mathbf{E}_{1}\{A\}}\bigg|^2\bigg\}+
3\mathbf{E}\bigg\{\bigg|\frac{B^{(c)}A^\circ_1}{A\mathbf{E}_{1}\{A\}}\bigg|^2\bigg\}.
\end{align*}
Averaging with respect to $\{a_{1i}\}$ and then using the H\"{o}lder inequality, we get
\begin{align*}&\mathbf{E}_1\bigg\{\bigg|\frac{B^{(c)\circ}_1}{\mathbf{E}_{1}\{A\}}\bigg|^2\bigg\}\le
C\frac{n^{-1}\mathrm{Tr\,}|G^{(1)}|^4e^{2c\mathcal{A}^{(1)}}}{p_n|\mathbf{E}_{1}\{A\}|^2}\le
C\frac{n^{-1}\mathrm{Tr\,}|G^{(1)}|^2e^{2c\mathcal{A}^{(1)}}}{p_n|\Im z|^2|\mathbf{E}_{1}\{A\}|^2}\\
&\le C\frac{(n^{-1}\mathrm{Tr\,}|G^{(1)}|^2)^{1-\varepsilon}
(n^{-1}\mathrm{Tr\,}|G^{(1)}|^2e^{2c\mathcal{A}^{(1)}/\varepsilon})^{\varepsilon}} {p_n|\Im
z|^2|\mathbf{E}_{1}\{A\}|^2}\le \frac{C(n^{-1}\mathrm{Tr\,} e^{2c\mathcal{A}^{(1)}/\varepsilon})^\varepsilon}{p_n|\Im
z|^{3+\varepsilon}|\mathbf{E}_{1}\{A\}|^{1+\varepsilon}},
\end{align*}
where $|G^{(1)}|^2=G^{(1)*}G^{(1)}$. Similarly, using that $\Im A=\Im z(1+(|G^{(1)}|^2a^{(1)},a^{(1)}))$ (see
(\ref{sp_rel})), we obtain
\begin{align*}&\mathbf{E}_{1}\bigg\{\bigg|\frac{B^{(c)}A^\circ_1}{A\mathbf{E}_{1}\{A\}}\bigg|^2\bigg\}
\le\mathbf{E}_{1}\bigg\{\bigg(\frac{(|G^{(1)}|^2a^{(1)},a^{(1)})^{1-\varepsilon}
(e^{2c\mathcal{A}^{(1)}/\varepsilon}|G^{(1)}|^2a^{(1)},a^{(1)})^{\varepsilon}|A^\circ_1|}{|\Im
z|(1+(|G^{(1)}|^2a^{(1)},a^{(1)}))|\mathbf{E}_{1}\{A\}|}\bigg)^2\bigg\}\\
&\le\frac{\mathbf{E}_{1}\big\{(e^{2c\mathcal{A}^{(1)}/\varepsilon}a^{(1)},a^{(1)})^{2\varepsilon}|A^\circ_1|^2\big\}}{|\Im
z|^{2}|\mathbf{E}_{1}\{A\}|^2} \le
 \frac{\mathbf{E}_{1}^{1-2\varepsilon}\{|A^\circ_1|^2\}
 \mathbf{E}_{1}^{2\varepsilon}\{(e^{2c\mathcal{A}^{(1)}/\varepsilon}a^{(1)},a^{(1)})|A^\circ_1|^2\}}
{|\Im z|^{2}|\mathbf{E}_{1}\{A\}|^2}\\
&\le C\frac{(n^{-1}\mathrm{Tr\,} |G^{(1)}|^2)^{1-2\varepsilon}(n^{-1}||G^{(1)}||^2\mathrm{Tr\,}
e^{2c\mathcal{A}^{(1)}/\varepsilon})^{2\varepsilon}} {p_n|\Im z|^{2}|\mathbf{E}_{1}\{A\}|^2}\le
C\frac{(n^{-1}\mathrm{Tr\,} e^{2c\mathcal{A}^{(1)}/\varepsilon})^{2\varepsilon}} {p_n|\Im
z|^{3+2\varepsilon}|\mathbf{E}_{1}\{A\}|^{1+2\varepsilon}}.
\end{align*}
The terms with $II$ and $III$ can be estimated similarly, if we use also the bound
\begin{equation}\label{l4.2}
  \mathbf{E}_{1}\big\{ \big((e^{c\mathcal{A}})_{11}
   -\mathbf{E}_{1}\big\{ (e^{c\mathcal{A}})_{11}\big\}\big)^2\big\}\le Cp_n^{-1}
   \mathbf{E}_{1}^{1/2}\big\{ (e^{8|c|\mathcal{A}})_{11}\big\}.
\end{equation}
To prove (\ref{l4.2}), we prove first that
\begin{equation}\label{l4.3}
\mathbf{E}_1\{((\mathcal{A}^{m})_{11}- \mathbf{E}_1\{(\mathcal{A}^{m})_{11}\})^2\}\le
Cm^22^m\mathbf{E}_1^{1/2}\{(\mathcal{A}^{4m-2})_{11}\}/p_n.
\end{equation}
It is easy to see that
\begin{equation}\label{l4.4}(\mathcal{A}^{m})_{11}=\sum_{k=1}^{[m/2]}\sum_{l_1+\dots+l_k=m-2k}
\Sigma^{(l_1)}\dots\Sigma^{(l_k)},\quad\Sigma^{(l_p)}:= (\mathcal{A}^{(1)l_p}a^{(1)},a^{(1)})
\end{equation}
 with
$\mathcal{A}^{(1)}$ and $a^{(1)}$ of (\ref{A^1}). Thus, using that for all $l$ $\mathbf{E}\{a_{ij}^l\}\ge 0$, we have
\begin{align*}
&\mathbf{E}_1\Big\{\Big(\Sigma^{(l_1)}\dots\Sigma^{(l_k)}-
\mathbf{E}_1\{\Sigma^{(l_1)}\}\dots\mathbf{E}_1\{\Sigma^{(l_k)}\}\Big)^2\Big\}\\
&\le k\sum_{j=1}^k \mathbf{E}_1^{1/2}\Big\{\Big(\Sigma^{(l_j)}-\mathbf{E}_1\{\Sigma^{(l_j)}\}\Big)^4\Big\}
\mathbf{E}_1^{1/2}\Big\{\prod_{i\not=j}(\Sigma^{(l_i)})^4\Big\}\\&\le Ckp_n^{-1}\sum_{j=1}^k
\mathbf{E}_1^{1/2}\{\Sigma^{(4l_j)}\}\mathbf{E}_1^{1/2}\Big\{\prod_{i\not=j}\Sigma^{(4l_i)}\Big\}
\le Ck^2p_n^{-1}
\mathbf{E}_1^{1/2}\Big\{\prod_{i}\Sigma^{(4l_i)}\Big\}.
\end{align*}
Taking the sum as in (\ref{l4.4}) and using the Schwarz inequality, we obtain (\ref{l4.3}). The Taylor expansion, the
Schwarz inequality, and (\ref{l4.3}) imply (\ref{l4.2}):
\begin{align*}
    &\mathbf{E}\{(e^{ c\mathcal{A}})_{11}-\mathbf{E}_1\{(e^{ c\mathcal{A}})_{11}\})^2\}\le p_n^{-1} \sum_{m=1}^\infty
   \frac{|2c^2|^m}{(m!)^2} \mathbf{E}\Big\{\Big((\mathcal{A}^{m})_{11}- \mathbf{E}_1\{(\mathcal{A}^{m})_{11}\}\Big)^2\Big\}\\&
   \le Cp_n^{-1} \sum_{m=1}^\infty\frac{m^2|2c|^{2m}}{(2m)!}\mathbf{E}_1^{1/2}\{(\mathcal{A}^{4m-2})_{11}\}\le
Cp_n^{-1} \mathbf{E}_1^{1/2}\bigg\{\sum_{m=1}^\infty\frac{|8c|^{4m}}{(4m)!}(\mathcal{A}^{4m-2})_{11}\bigg\}\\&\le
Cp_n^{-1}\mathbf{E}_1^{1/2}\bigg\{(e^{8|c|\mathcal{A}})_{11}\bigg\}.
\end{align*}
 Lemma \ref{l:4} is proven. $\square$

 The next step is the analog of Proposition \ref{p:joh}
 \begin{proposition}\label{p:joh1} For any $\varphi\in\mathcal{H}_s^{(c)}$
\begin{equation}\label{pj1.1}
\mathbf{Var}\{\mathcal{N}_n[\varphi ]\}\le C_s||\varphi^{(c)}||_s^2\int_0^\infty dy
e^{-y}y^{2s-1}\int_{-\infty}^\infty\mathbf{Var}\{\gamma_n^{(c)}(x+iy)\}dx,
\end{equation}
where $ \widetilde\varphi(\lambda)=\varphi(\lambda)\cosh^{-1}(c\lambda)$ and $||\widetilde\varphi||_s$ is defined in
(\ref{norm}).
\end{proposition}
 The proof of Proposition \ref{p:joh1} coincides with that of Proposition \ref{p:joh}, if we replace
  the operator $\mathcal{V}$ of (\ref{calV}) by
the  operator $\mathcal{V}^{(c)}$ whose Fourier transform has the kernel
\[\widehat{\mathcal{V}^{(c)}}(k_1,k_2)=\mathbf{Cov}\{\mathrm{Tr}\cosh(c\mathcal{A})
e^{ik_1\mathcal{A}},\mathrm{Tr}\cosh(c\mathcal{A})e^{ik_2\mathcal{A}}\},\quad
\mathbf{Var}\{\mathcal{N}_n[\varphi]\}=(\mathcal{V}^{(c)}\widetilde\varphi,\widetilde\varphi).
\]

Now  one can derive (\ref{main_1}) from Lemma \ref{l:4} and Proposition \ref{p:joh1} by the same argument that we used
in Lemma \ref{l:2} to derive (\ref{l2.1}) from Lemma \ref{l:1} and Proposition \ref{p:joh}.

Having in mind the bound (\ref{main_1}), we can derive Theorem \ref{t:CLT2} from Proposition \ref{p:CLTcont}, if we are
able to prove CLT for some dense subset of $\mathcal{H}_s^{(c)}$, e.g., for  $\varphi$ with finite supports. But if
$\varphi$ has a finite support and belongs to $\mathcal{H}_s^{(c)}$, it belongs also automatically to $\mathcal{H}_s$,
thus we can apply Theorem \ref{t:CLT} to it. This completes the proof of Theorem \ref{t:CLT2}.

\end{document}